\documentclass[%
 reprint,
superscriptaddress,
 amsmath,amssymb,
 aps,
]{revtex4-2}

\usepackage{graphicx}
\usepackage{dcolumn}
\usepackage{bm}
\usepackage{hyperref}
\usepackage[hyphenbreaks]{breakurl}

\UseRawInputEncoding
\usepackage[section]{placeins}

\begin{document}

\preprint{APS/123-QED}

\title{High-dimensional quantum key distribution using energy-time entanglement\\over 242 km partially deployed fiber}

\author{Jingyuan Liu}
\author{Zhihao Lin}
\author{Dongning Liu}
\author{Xue Feng}
\author{Fang Liu}
\author{Kaiyu Cui}
\affiliation{Frontier Science Center for Quantum Information, Beijing National Research Center for Information Science and Technology (BNRist), Electronic Engineering Department, Tsinghua University, Beijing 100084, China.}

\author{Yidong Huang}
\author{Wei Zhang}
\email{zwei@tsinghua.edu.cn}
\affiliation{Frontier Science Center for Quantum Information, Beijing National Research Center for Information Science and Technology (BNRist), Electronic Engineering Department, Tsinghua University, Beijing 100084, China.}
\affiliation{Beijing Academy of Quantum Information Sciences, Beijing 100193, China.}

\begin{abstract}
Entanglement-based quantum key distribution (QKD) is an essential ingredient in quantum communication, owing to the property of source-independent security and the potential on constructing large-scale quantum communication networks. However, implementation of entanglement-based QKD over long-distance optical fiber links is still challenging, especially over deployed fibers. In this work, we report an experimental QKD using energy-time entangled photon pairs that transmit over optical fibers of 242 km (including a section of 19 km deployed fibers). High-quality entanglement distribution is verified by Franson-type interference with raw fringe visibilities of 94.1$\pm$1.9\% and 92.4$\pm$5.4\% in two non-orthogonal bases. The QKD is realized through the protocol of dispersive-optics QKD. A high-dimensional encoding is applied to utilize coincidence counts more efficiently. Using reliable, high-accuracy time synchronization technology, the system operates continuously for more than 7 days, even without active polarization or phase calibration. We ultimately generate secure keys with secure key rates of 0.22 bps and 0.06 bps in asymptotic and finite-size regime, respectively. This system is compatible with existing telecommunication infrastructures, showing great potential on realizing large-scale quantum communication networks in future.
\end{abstract}

\maketitle


\section{\label{sec:level1}Introduction}
Quantum key distribution (QKD) establishes information-theoretically secure keys between two remote users. Since the first QKD protocol (BB84) was proposed\cite{bennett1984}, many progresses have been achieved theoretically and experimentally in tackling the potential attack from eavesdropper\cite{lo2005,wang2005,pironio2009,lo2012,yin2016,zhang2022}, extending the QKD distance\cite{scheidl2009,lucamarini2018,yin2020,chen2021,wang2022,zeng2022} and improving the secure key rate\cite{ali2007,pseiner2021}. In the experimental realization of QKDs, there are mainly two schemes, namely the prepare-and-measure scheme\cite{bennett1984,bennett19921} and the entanglement-based scheme\cite{ekert1991,bennett19922}. The entanglement-based QKD exploits the intrinsic correlation of quantum entanglement to distribute cryptographic keys, which dispenses with the need of random number generators. It also has source-independent security and possibilities to realize device-independent QKD\cite{pironio2009,zhang2022}. Moreover, researchers have recently shown that through entanglement distribution between end users, large-scale quantum networks with fully-connected topology can be established\cite{wengerowsky2018,liu2022}. It can be expected that the long-distance entanglement-based QKD with deployed fiber is a crucial technology for future global quantum communication networks. 

Enormous efforts have been made to realize entanglement-based QKD under long-distance fiber transmission in laboratory\cite{honjo2008,takesue2010} and in field\cite{wengerowsky2020}. Recently, researchers have shown the entanglement distribution of polarization entangled photon pairs along 248 km deployed fiber and calculated its secure key rates\cite{neumann2022}. Since distribution of polarization entanglement over optical fibers is sensitive to environment variations, dedicated polarization calibration is necessary in these works to compensate for polarization drift, and the active calibration process is required every few hours to reduce the QBER to a sufficiently low level\cite{neumann2022,treiber2009}. Among entanglement on different degrees of freedom (DOFs), the energy-time entanglement and its discrete form, i.e., time-bin entanglement, are promising candidates for long-distance entanglement-based QKD thanks to its convenience in generation and manipulation, and resilience to environment variations in fiber transmission. To the best of our knowledge, the maximum fiber transmission distance of entanglement distribution of energy-time and time-bin entanglement is 300 km in laboratory\cite{inagaki2013}, and about tens of kilometers in field\cite{tittel1998,salart2008}. As for entanglement-based QKD using energy-time and time-bin entanglement, researches are mainly carried out in the laboratory environment\cite{honjo2008,takesue2010}. As far as we know, the longest implementation with deployed fiber is demonstrated in a link of a quantum network, with a total fiber length of 108 km including deployed fiber of 26.8 km\cite{fitzke2022}. The experiment was conducted for about 12 h with the aid of time synchronization and phase alignment. It makes the real-scenario implementation of entanglement-based QKD with longer transmission and long-term operation become a critical issue to be solved.

Quantum entanglement is the precious resource for entanglement-based QKD. To exploit detected photon-pairs (or photons) more efficiently, high-dimensional QKD (HD-QKD) is proposed to perform multi-level encoding using qudits with dimension D\textgreater2\cite{thew2004}. The HD-QKD can increase the shared secure keys between Alice and Bob for each detected photon-pair (or photon), meanwhile, it would offer better robustness against channel noise. High-dimensional quantum information can be encoded in many DOFs, such as energy-time\cite{ali2007,thew2004}, transverse momentum\cite{etcheverry2013}, orbital angular momentum\cite{molina2004,mafu2013}. Recently, many progresses have been made in dispersive-optics QKD (DO-QKD)\cite{mower2013,lee2014,lee2019,liu2019}, which encodes photons in continuous temporal modes with a large-alphabet way, combing the high-dimensional Hilbert spaces of continuous-variable QKD (CV-QKD) with single-photon detection. The DO-QKD protocol uses dispersive optics to construct frequency bases, avoiding the need of phase stabilization in Franson interferometers used in regular time-bin configuration. 

In this work, we report an experimental demonstration of entanglement-based DO-QKD over fiber transmission of 242 km, including a 19 km deployed fiber. Time synchronization based on optical pulses are implemented in the system. We perform high-quality entanglement distribution verified by Franson-type interference with high visibility. Our system shows excellent stability and robustness over 7 days of continuous operation, even without active polarization or phase calibration. In key generation, a high-dimensional encoding format is utilized to increase the number of secure keys extracted from each detected coincidence count. Then the security analysis using time-frequency covariance matrix is implemented, showing protocol security against Gaussian collective attacks. Error correction and privacy amplification are finally carried out to generate secure keys in asymptotic and finite-size regime. As far as we know, it is the implementation with longest distance of fiber-based entanglement-based QKD field test using energy-time entanglement.

\section{Results}

\subsection{\label{sec:level2}Experimental setup}
\begin{figure*}
	\includegraphics[width=\linewidth]{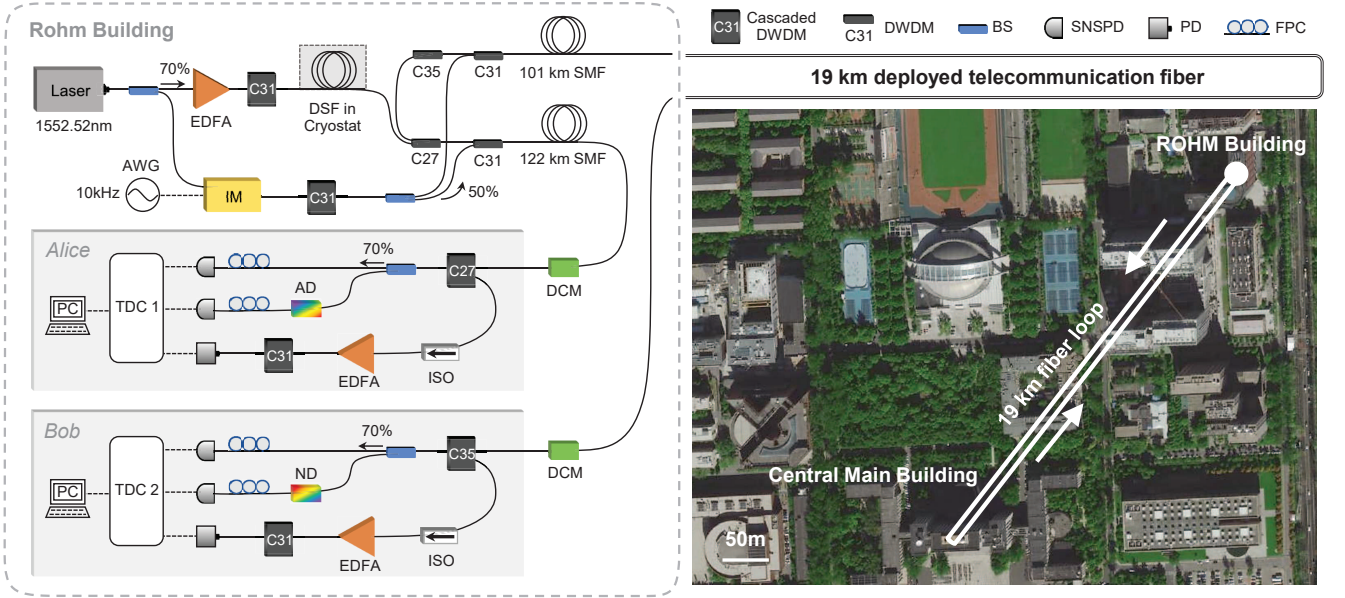}
	\caption{\label{fig1}Schematics of the experimental setup. The source, users and fiber spools are set at the ROHM Building while a deployed fiber loop is laid between the ROHM Building and the Central Main Building in the campus of the Tsinghua University. The source generates entangled photon pairs and light pulses used for synchronization between two users. After transmission, Alice and Bob perform wavelength demultiplexing and singe-photon detection in an asymmetric basis selection configuration. The arrival time data are recorded using different TDCs and then the time calibration is implemented in post-processing. EDFA: Erbium-doped fiber amplifier; BS: beam splitter; ISO: isolator; AD: anomalous dispersion; ND: normal dispersion; FPC: fiber polarization controller; PC: personal computer. Imagery \copyright 2022 AutoNavi, DigitalGlobe \& spaceview. Map data \copyright 2022 AutoNavi.}
\end{figure*}

We implement entanglement-based DO-QKD with source-in-the-middle configuration. The illustration of experimental setup is schematically depicted in Fig.~\ref{fig1}. The pump laser runs in continuous-wave (CW) mode with central wavelength of 1552.52 nm, which corresponds to 100 GHz International Telecommunication Union (ITU) channel of C31. A beam splitter separates the pump light into two parts, one of which pumps a 280 m dispersion shifted fiber (DSF) to generate energy-time entangled photon pairs through spontaneous four-wave mixing (SFWM) effect. Then dense wavelength division multiplexers (DWDMs) separate signal and idler photons into 100 GHz ITU channels of C35 (1549.32 nm) and C27 (1555.75 nm), respectively. The other part of the pump light incidents an intensity modulator (IM) to generate synchronization pulses. The IM is driven by a rectangular microwave signal generated by arbitrary waveform generator (AWG) with 10 kHz repetition frequency and 10 ns pulse width. 

To track time fluctuation in transmission fibers, the generated signal and idler photons are multiplexed with synchronization pulses, then sent to Bob and Alice through quantum channels, respectively. The quantum channels have a total length of 242 km, composed of 223 km single mode fiber (SMF) spools and a 19 km deployed SMF link. The deployed fiber link includes 12 pieces of dark fibers in a fiber cable between two buildings in the campus of Tsinghua University (the ROHM Building and the Central Main Building of Tsinghua University, as shown in Fig.~\ref{fig1}), which were spliced end-to-end. The quantum channel from the source to Alice composes of 122 km SMF spools with a loss of 24 dB, while that from the source to Bob composes of 101 km SMF spools and the 19 km deployed SMF loop with a total loss of 27 dB. The SMFs used in the experiment are standard telecommunication G.652 fibers with chromatic dispersion of 17 ps/(km$\cdot$nm) at 1550 nm. The chromatic dispersion of transmission fibers is precisely compensated using dispersion compensation modules (DCMs). 

At each user's end, the synchronization pulses are filtered out by cascaded DWDM components, amplified and then detected by photodetectors (PDs). For single-photon detection, Alice and Bob use 70:30 beam splitters to randomly choose between the time and frequency bases. This asymmetric basis selection configuration can effectively increase raw key rate, while it allows to accumulate enough coincidences for security analysis\cite{lo2004}. In time bases, the arrival times of photons are directly detected by superconducting nanowire single-photon detectors (SNSPD). While in frequency bases, the arrival times of photons are detected after they passing through a dispersion module with anomalous (normal) dispersion at Alice's (Bob's) side. The nonlocal dispersion is cancellated by the intrinsic nature of energy-time entanglement when two photons of an entangled photon pair are both detected in frequency bases\cite{franson1992}. The electric pulses output from PDs and SNSPDs are then recorded by the time-to-digital convertors (TDCs). All the recorded arrival times of photons at Alice's and Bob's sides are rescaled in post-processing based on the synchronization pulses, and the calibrated arrival times are utilized in key generation and security analysis.

\subsection{DO-QKD Protocol}
In order to fully exploit every recorded coincidence event, a high-dimensional encoding format is utilized in key generation. Alice and Bob sift their keys in a three-level bin sifting process\cite{liu2019}. The time streams of two users are divided into consecutive time frames, time slots and time bins. A frame contains $M=2^{D}$ slots and a slot contains $I$ bins. Each user selects the frames which contains only one single-photon event and discards other frames. Then they communicate the frame number and bin number of every retained single-photon event in each side through classical channel. Only photons from two users has the same frame number and the same bin number would be registered as a successful coincidence event. This strategy effectively decreases the quantum bit error rate (QBER) caused by timing jitters of detectors and electronic devices. Then the raw keys are generated by the slot numbers of the single-photon events.

The security analysis of DO-QKD is based on treatment of single-photon events and the well-established proofs of Gaussian CV-QKD\cite{serafini2004,garcia2006}. By calculating time-frequency covariance matrix between photons' arrival times of Alice and Bob, the security of protocol against Eve's Gaussian collective attacks has been proven\cite{mower2013}. The secure key capacity represents the number of secure keys that can be extracted from each coincidence count. It is denoted as
\begin{equation}
	\Delta I=\beta I(A;B)-\chi (A;E)-\Delta _{FK},
\end{equation}
where $\beta$ is the reconciliation efficiency, $I(A;B)$ is Shannon mutual information between Alice and Bob, $\chi (A;E)$ is Eve's Holevo information, and $\Delta _{FK}$ accounts for penalty of finite-size effect.

When taking the finite length of generated keys into consideration, each stage of the QKD protocol has a probability to fail\cite{lee2015}. The tolerated failure probability of the whole protocol $\varepsilon_{s}=\varepsilon_{ver}+\varepsilon_{PA}+n_{PE}\varepsilon_{PE}+\overline{\varepsilon}$, which is the sum of failure probabilities in stages of error correction and verification, privacy amplification, parameter estimation and estimating the smooth min-entropy. In the process of information reconciliation, we combine symmetric blind information reconciliation\cite{kiktenko2017}, which is based on low-density parity-check codes, with layered scheme\cite{zhou2013} suitable for large-alphabet QKD to further improve the reconciliation efficiency. Then hash-function-based verification\cite{fedorov2018} is introduced to reduce the QBER after information reconciliation. Finally, to extract secure keys from reconciled keys, privacy amplification is implemented using Toeplitz matrix\cite{bennett1995}.

\subsection{Entanglement distribution}
\begin{figure*}
	\includegraphics{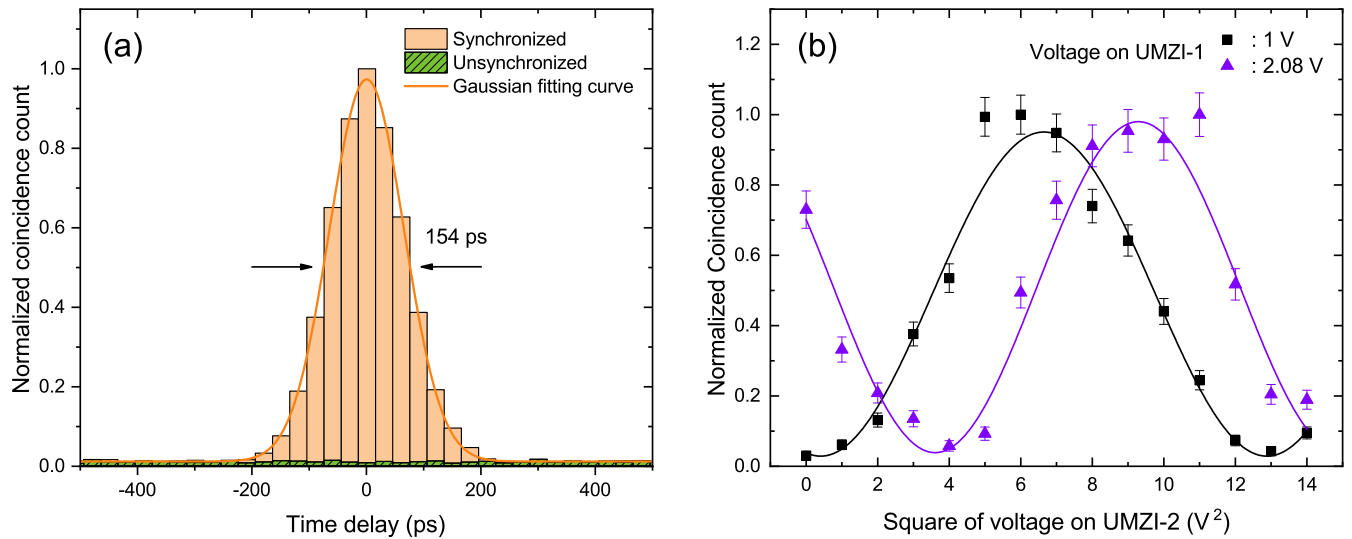}
	\caption{\label{fig2}Characterization of energy-time entanglement after 242 km fiber transmission. (a) Typical two-photon coincidence measurement. Orange (Green) bars represent the coincidence between Alice and Bob with (without) time synchronization algorithm performed in post-processing, with respect to the same set of single-photon arrival time data. Coincidence counts in each time bin are normalized to the maximum coincidence count. The Gaussian fitting curve of the coincidence peak with time synchronization is shown in orange line with FWHM of 154 ps. Bin width for coincidence measurement here is 30 ps. (b) Franson-type interference fringes without subtracting the accidental coincidence counts. The black squares and purple triangles are normalized coincidence counts when the voltage on UMZI-1 is 1 V and 2.08 V. The acquisition time is 60 min for each point and the error bars are obtained by Poissonian photon-counting statistics. The solid lines are sinusoidal fitting curves of the experimental data.}
\end{figure*}

The high-quality entanglement distribution of energy-time entangled photon states is demonstrated with 242 km transmission fiber. First of all, we perform coincidence measurement between signal and idler photons received by two users with a slightly different experimental setup from that in Fig.~\ref{fig1}. After being separated from synchronization pulses, the signal and idler photons are directly detected by SNSPDs and their arrival times are recorded by TDCs at each user's end. Fig.~\ref{fig2}(a) shows the typical coincidence measurement after transmission with normalized amplitude. The coincidence measurement with calibrated arrival times of photons after time synchronization is given in orange bars. It reveals that the temporal correlation between entangled photon pairs is well preserved after transmission to distant synchronized users. A gaussian fitting is implemented on the coincidence peak, shown in orange line. The full width at half maximum (FWHM) of the coincidence peak is 154 ps, which mainly attributes to the timing jitters of detectors and electronic circuits. Strict temporal filtering is applied to reduce the accidental coincidence counts caused by noise photons. Under a coincidence window of 240 ps, covering most of the coincidence peak, the coincidence count rate is 0.59 cps with coincidence to accidental coincidence rate (CAR) of 64.9. For comparison, we also present coincidence measurement result with the raw arrival time data without time synchronization, given in green bars. In this case, instead of accumulated coincidence peak, the coincidences are full of accidental events with equal probability in every time bin. The temporal correlation is undetectable due to the asynchronous clocks of two TDCs and the time fluctuation caused by long-distance fibers. The following tests of the paper are implemented with time synchronization algorithm performed in post-processing.

To evaluate the quality of the energy-time entanglement after long-distance transmission, we perform Franson-type interference\cite{franson1989}. The signal and idler photons are detected and recorded after they passing through two unbalanced Mach-Zehnder interferometers (UMZI) placed at each user's side, UMZI-1 and UMZI-2, respectively. The unbalanced time $\Delta T$ between the two arms of the UMZI is about 400 ps, which is much larger than the single-photon coherence time of about6 ps to avoid single-photon interference. By applying voltage on one arm of a UMZI, we can change the phase difference between its two arms. The voltage of the UMZI-1 is set to be 1 V and 2.08 V, constructing two non-orthogonal bases with phase difference of $\pi/2$, while scanning the voltage of UMZI-2 to obtain interference fringes. We select the central coincidence peak with a coincidence window of 250 ps and the interference results are shown in Fig.~\ref{fig2}(b). The coincidence counts in two sets are normalized to the maximum values of each set. It can be seen that after transmission of 242 km, the visibilities of two-photon interference are 94.1$\pm$1.9\% and 92.4$\pm$5.4\% in two non-orthogonal bases, without subtracting the accidental coincidence counts. We calculate the S parameter to be 2.61$\pm$0.15, with a violation of the Clauser-Horne-Shimony-Holt (CHSH) Bell inequality S$\leqslant$2 by 4 deviations\cite{clauser1969}. The result of Franson-type interference in back-to-back configuration is given in Appendix~\ref{app1}. Our results confirm that the energy-time entanglement is well-preserved after long-distance entanglement distribution.

\subsection{Quantum key distribution}
Using the distributed energy-time entangled photon pairs, the entanglement-based quantum key distribution is demonstrated. In the QKD setting, the time bases and frequency bases are constructed at each user's end, as depicted in Fig.~\ref{fig1}. Alice and Bob use all the coincidence events which they both detected in frequency bases to perform parameter estimation. With respect to coincidence events which they both detected in time bases, they take a part of them for parameter estimation. The amount of this part is equal with those detected in frequency bases. The remained coincidence events are used for key generation. Before the QKD starts, Alice and Bob need to optimize and determine the time-encoding parameters for key generation. A higher encoding dimension or a larger time bin width would result in a higher raw key rate, however, it may also lead to a higher QBER. The value of QBER affects the length of reconciled keys after error correction so that it is also crucial to the secure key rate (SKR). Hence, trade-off between raw key rate and QBER should be carefully made in parameter optimization. In our experiment, the QBER is set to below an upper bound of 8\% and the encoding parameters are optimized accordingly to maximize the key rate.

\begin{figure*}
	\includegraphics[width=13.5cm]{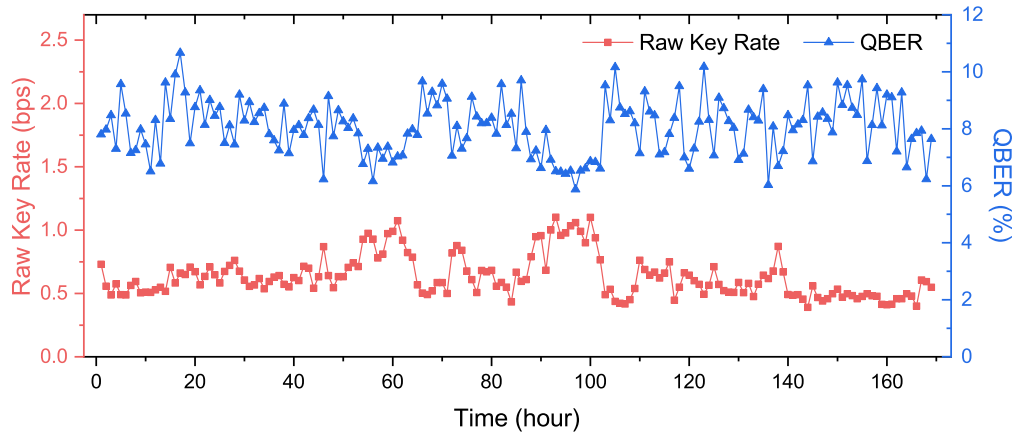}
	\caption{\label{fig3}System stability with transmission distance of 242 km during 169-hour data collection. The raw key rate and QBER versus time are shown in rosy squares and blue triangles, respectively. Each point represents the average value over 1 h of data.}
\end{figure*}

The 242 km DO-QKD system operates continuously for an acquisition time of 169 hours, demonstrating the stability and robustness of the QKD link without active phase or polarization calibration. All the arrival times of single-photon events are recorded and calibrated with time synchronization algorithm by Alice and Bob. Then, the raw key generation is performed on data of every hour with time-encoding parameters of $D=3, I=3$ and bin width = 240 ps. Hourly raw key rates and QBERs versus time are shown in Fig.~\ref{fig3}. The raw key rate distribution is concentrated around 0.63 bps with a standard deviation of 0.16 bps, and the corresponding QBER values have a mean value is 8.04\% with a standard deviation of 0.98\%. The fluctuation of raw key rates is mainly caused by fluctuation of SNSPD's detection efficiency when polarization slowly drifts in the long fiber. The QBER is stable in general with statistical fluctuations during the test. After 169 hours of data acquisition, we obtain raw keys with a total number of 385, 665 bits. 

After the raw keys are generated, efficient information reconciliation is performed to eliminate error bits between the two users. In the case with 242 km fiber transmission, we experimentally achieve a reconciliation efficiency of 89.7\% by combining symmetric blind information reconciliation method\cite{kiktenko2017} with layered scheme\cite{zhou2013}. The tolerated failure probability $\varepsilon_{s}$ is set as $10^{-5}$ in the experiment, and failure probability of error correction and verification $\varepsilon_{ver}$ is $2\times10^{-12}$. By calculating time-frequency covariance matrix and excess noise factor introduced by quantum channel and potential eavesdropper\cite{mower2013}, the Shannon mutual information between Alice and Bob $I(A;B)$ is determined as 1.87 bpc, and Eve's maximum accessible information $\chi(A;E)$ is 0.65 bpc in asymptotic regime. When considering the finite-size effect with the number of total coincidence counts $N$ of $\sim2\times10^{5}$ between Alice and Bob, the $\chi(A;E)$ increases to 1.02 bpc while an additional subtraction $\Delta_{FK}$ of 0.39 bpc should be considered. The secure key capacity is 1.02 bpc and 0.27 bpc in asymptotic and finite-size regime, while the corresponding SKR is 0.22 bps and 0.06 bps, respectively. Privacy amplification based on Toeplitz matrix further extracts finite-size secure keys of 34, 905 bits. 

\begin{figure}[b]
	\includegraphics[width=8cm]{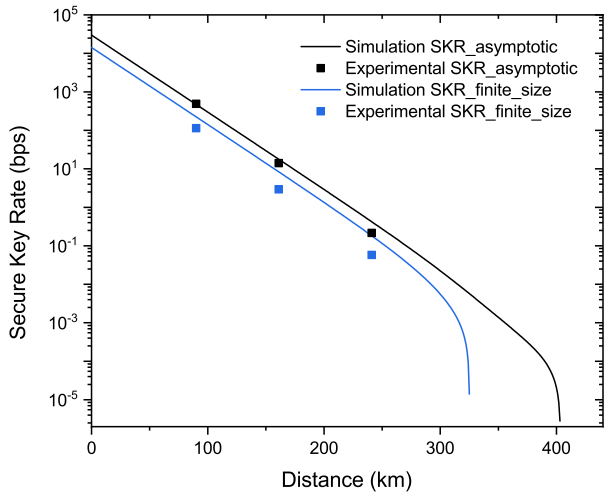}
	\caption{\label{fig4} Secure key rates of the DO-QKD simulation and experiment. SKR values are plotted versus distance of SMF. The simulations of DO-QKD performance in asymptotic and finite-size regime are shown in black and blue curves, respectively. Corresponding experimental results are given in black and blue squares. The number of total coincidence events $N$ in simulation and experiment is $\sim2\times10^{5}$. The sum of dark count rate of SNSPD and the noise photon rate introduced by synchronization pulses is $\sim$300 Hz.}
\end{figure}

Figure~\ref{fig4} shows the secure key rates of DO-QKD under different transmission distances. In experiment, we conduct DO-QKD tests under 90 km, 161 km and 242 km by changing the length of fiber spools while the 19 km deployed fiber loop is included in the system throughout the experiments. In each case, the chromatic dispersion of transmission fibers is compensated by dispersion compensation modules, ensuring that the temporal correlation is well-preserved after transmission. The parameters for key generation are optimized case by case to maximize key rates (see details in Appendix~\ref{app2}). To accumulate enough coincidence events for finite-size analysis, the acquisition times of the three cases are 5 min, 3.5 h and 169 h, respectively, and the number of total coincidence events $N$ are $\sim2\times10^{5}$ in all cases. The experimental results are shown in Fig.~\ref{fig4}, in which the asymptotic and finite-size SKRs are shown in black and blue squares, respectively. Moreover, we simulate the expected performance of the DO-QKD system based on a previously proposed theoretical model\cite{liu2021}, considering the process of the quantum light source, transmission, single-photon detection and post-processing. The simulation model of DO-QKD is based on the assumptions that if one photon in an entangled photon pair is detected at a specific time at Alice's side, the detection time of the other photon in the pair at Bob's side would have a Gaussian distribution, and noise photons always have a uniform distribution in a time frame. The parameters used in simulation are the averaged values of experimental parameters of three cases. The simulation curves are also shown in Fig.~\ref{fig4}. It shows that the experimental results are in good agreement with those of the simulation. However, the difference between asymptotic and finite-size regime in simulation is smaller than that in experiment, which suggests that the Shannon mutual information is overestimated in simulation because of a rather ideal statistical distribution assumption. Our result reveals that the entanglement-based DO-QKD scheme shown in Fig.~\ref{fig1} can tolerate transmission of 400 km in asymptotic regime, and about 300 km in finite-size regime.

\section{Discussion}
By using high-accuracy optical time synchronization method, our QKD system using energy-time entanglement exhibits excellent stability over a long time period and system robustness against channel fluctuation. In the experiment over 7 days, the system suffered from several transient perturbations manifested by missing several synchronization pulses which may be caused by unstable electrical environment. The impact of these perturbations can be removed in post-processing by discarding the single-photon events in these transient periods and maintaining synchronization after the perturbations. Notably, there is no active phase or polarization calibration implemented in our experiment, while it still shows long-term stability over 7 days with fully passive experimental setup. 

We have shown that our system can tolerate more than 51 dB transmission loss with distance over 242 km. The corresponding finite-size SKR is at the order of $10^{-2}$ bps, which is at a relatively low level in our experiment. At the source side, we choose a moderate pump power to mitigate the multi-pair emission in a time frame. The coincidence count rate is restricted in this situation so that an efficient encoding method becomes rather necessary. We use a high-dimensional encoding format to maximize the use of coincidence counts in key generation. In our work, the encoding dimension $D$ is 3, and we obtain a secure key capacity of 1.02 bpc in asymptotic regime and 0.27 bpc in finite-size regime. A comparison between our work with previous entanglement-based QKD works using energy-time and time-bin entanglement is shown in Table~\ref{table1}. There are several possible improvements of our QKD system. Firstly, the generation rate of entangled photon pairs still has space to be increased, under the limit that the possibility of multi-pair emission in a time frame is sufficient low. Secondly, the performance of QKD would improve considerably if we use advanced commercial SNSPD system with detection efficiency over 90\% and timing jitter lower than 50 ps. Thirdly, by employing a longer acquisition time, the finite-size SKR would closely approaches to the asymptotic SKR when increasing the total number of coincidences $N$ between two users\cite{lee2015}. We estimate the potential performance of the DO-QKD system with the above improvements. Through theoretical calculation, it can be expected that the finite-key SKR under 242 km transmission would be improved to the order of 1 bps using a periodically poled LiNbO$_{3}$ source described in Ref.~[\onlinecite{zhang2021}], offering higher generation rate of entangled photon pairs and higher CAR (see details in Appendix~\ref{app2}).

\begin{table*}
	\caption{\label{table1}%
		Comparison with other entanglement-based QKD using energy-time and time-bin entanglement.
	}
	\begin{ruledtabular}
		\begin{tabular}{cccccc}
			&Length (km)&Fiber type&Protocol
			& Asymptotic SKR (bps) &Finite-size SKR (bps)\\
			\hline
			T. Honjo, et al. (2008)\cite{honjo2008}& 100 & Laboratory DSF & BBM92 & 0.14 & --  \\
			T. Zhong, et al. (2015)\cite{zhong2015}& 20 & Laboratory SMF & HD-QKD & -- & 2.7M \\
			E. Fitzke, et al. (2022)\cite{fitzke2022}]& 108 & Partially deployed SMF & BBM92 & 6.3 & -- \\
			This work& 242 & Partially deployed SMF & DO-QKD &0.22 & 0.06 \\
		\end{tabular}
	\end{ruledtabular}
\end{table*}

\section{Conclusion}
In summary, we have demonstrated a high-dimensional DO-QKD system with 242 km transmission fiber including deployed fiber of 19 km. The high-quality entanglement distribution is verified using Franson-type interference. The raw visibilities are 94.1$\pm$1.9\% and 92.4$\pm$5.4\% in two non-orthogonal bases, indicating the violation of CHSH Bell inequality by 4 deviations. Then the DO-QKD system continuously operates for over 7 days with time synchronization of the two users. We perform high-dimensional key generation, efficient information reconciliation and privacy amplification to generated secure keys with asymptotic SKR of 0.22 bps and finite-size SKR of 0.06 bps. The result shows outstanding stability and robustness of our system without active phase or polarization calibration. It has great potential on realizing entanglement-based QKD in large-scale optical fiber networks, paving the way towards future global quantum communication networks.

\begin{acknowledgments}
	This work has been supported by the National Key R\&D Program of China (2018YFB2200400), Natural Science Foundation of Beijing (Z180012), National Natural Science Foundation of China (61875101, 91750206), and Tsinghua Initiative Scientific Research Program.
	
	W. Z. and J. L. proposed the scheme. J. L., Z. L. and D. L. performed experiments. J. L. perform simulation and data analysis. Z. L. perform error correction and privacy amplification. W. Z., J. L. wrote the manuscript. Y. H. revised the manuscript and supervised the project. X. F., F. L., K. C. contributed to experiment design and the revision of the manuscript.
\end{acknowledgments}

\appendix

\section{\label{app1}Characterization of entanglement source}
In our work, we use a 280 m DSF as the nonlinear medium of the quantum light source based on the SFWM effect. The DSF is placed in a cryostat and cooled to $\sim$2.2 K to reduce the noise introduced by spontaneous Raman scattering. The pump power of the quantum light source is 19.4 dBm, producing about $8.8\times10^{6}$ entangled photon pairs per second. To test the quality of energy-time entanglement photon pairs output from the quantum light source, we perform Franson-type interference in a back-to-back configuration. The raw visibilities are 96.4$\pm$1.8\% and 94.6$\pm$0.6\% of the fringes in two non-orthogonal bases, without subtracting the accidental coincidence counts. The $S$ parameter is 2.68$\pm$0.02 with a violation of the CHSH Bell inequality by 34 deviations. It shows that our source generates high-quality energy-time entangled photon pairs.

\section{\label{app2}Secure key generation and simulation}
The DO-QKD performances with three different transmission distances are shown in the main text. The details of key generation in finite-size regime are shown in Table~\ref{table2}. The time encoding parameters are $N=3$ and $I=3$ for all cases. We maximize the time bin in each case to obtain a higher key rate when the QBER is below an upper bound of 8\%. In error correction, shortening and puncturing technique are used to optimize the reconciliation efficiency, by adding shortened and punctured bits in each block with a block length of 1944\cite{kiktenko2017,fedorov2018}. 

\begin{table}[b]
	\caption{\label{table2}%
		Details of finite-size key generation with different transmission distances.
	}
	\begin{ruledtabular}
		\begin{tabular}{cccc}
			Length (km)& 90 & 161 & 242
			\\
			\hline
			Acquisition time& 5 min & 3.5 h & 169 h\\
			Bin width(ps)& 350 & 250 & 240\\
			QBER(\%)& 7.92 & 7.94 & 7.90\\
			$\beta I(A;B)$(bpc)& 1.70 & 1.69 & 1.68\\
			$\chi(A;E)$(bpc)& 1.03 & 1.01 & 1.02\\
			$\Delta_{FK}$(bpc)& 0.39 & 0.39 & 0.39\\
			$\Delta I$(bpc)& 0.28 & 0.29 & 0.27\\
			SKR(bps)& 113.97 & 2.94 & 0.06\\
			Secure key(bit)& 34.0k & 36.7k & 34.9k\\
		\end{tabular}
	\end{ruledtabular}
\end{table}

To estimate the potential performance of the DO-QKD system, the quality of quantum light source in ref.~[\onlinecite{zhang2021}], which is based on second-order nonlinear process, is used for simulation. Through the given results in the reference, we estimate that a generation rate of entangled photon pairs at 13 MHz is accessible with corresponding CAR of about 200. In our simulation, the detection efficiency of SNSPDs is set to be 90\% and the timing jitter is 50 ps. The sum of dark count rate of SNSPD and the noise photon rate introduced by synchronization pulses is 200 Hz. The acquisition time is 169 h to perform finite-size analysis at 242 km transmission, which corresponds to $N$ of $3.5\times10^{6}$. Other parameters are the same as those in our experiment.

\bibliography{manuscript}

\end{document}